\documentstyle[twocolumn,aps]{revtex}

\draft
\begin{document}
 

\title{Field Induced Reduction of 
the Low Temperature Superfluid Density in YBa$_2$Cu$_3$O$_{6.95}$
[Physical Review Letters {\bf 83}, 4156 (1999)]} 

\author{J.E.~Sonier$^{1,2}$, J.H.~Brewer$^{2,3}$, R.F.~Kiefl$^{2,3}$, G.D.~Morris$^{2}$, 
R.I.~Miller$^{2,3}$, D.A.~Bonn$^{3}$,
J.~Chakhalian$^{2,3}$, R.H.~Heffner$^{1,2}$, W.N.~Hardy$^{3}$ and R.~Liang$^{3}$}
\address{$^1$Los Alamos National Laboratory, Los Alamos, New Mexico 87545, USA}
\address{$^2$TRIUMF, Vancouver, British Columbia, Canada, V6T 2A3} 
\address{$^3$Department of Physics and Astronomy, University of British Columbia,
Vancouver, British Columbia, Canada V6T 1Z1}

\date{ \rule{2.5in}{0pt} }

\maketitle 
\begin{abstract} \noindent
A novel high magnetic field (8~T) spectrometer for
muon spin rotation ($\mu$SR) has been used to measure the
temperature dependence of the in-plane
magnetic penetration depth $\lambda_{ab}$ in YBa$_2$Cu$_3$O$_{6.95}$. 
At low $H$ and low $T$, $\lambda_{ab}$ exhibits the characteristic linear
$T$-dependence associated with the energy gap of a $d_{x^2-y^2}$-wave superconductor.
However, at higher fields $\lambda_{ab}$ is essentially temperature 
independent at low $T$. We discuss
possible interpretations of this surprising new feature in the
low-energy excitation spectrum.
\end{abstract} 

In a superconductor, the resistance to the flow of electric current drops
to an unmeasurably small value below a certain critical temperature $T_c$. 
This remarkable characteristic is due to the formation of pairs of electrons 
(or holes), called ``Cooper pairs'', which link together and
carry the charge through the sample with virtually no opposition.
To break apart the pairs, an additional 
energy is needed to excite individual electrons above an energy
gap which exists at the Fermi surface in the superconducting state. The
nature of these elementary excitations, known as ``quasiparticles'' (QPs), 
is directly related to the size and symmetry of the energy gap. The gap itself
reflects the symmetry of the pair wave function (or order parameter),
knowledge of which is essential to understanding the physics of
the underlying mechanism responsible for superconductivity.  

A major breakthrough in the study of high-$T_c$ cuprate 
superconductors (HTSCs) came when
it was realized that the symmetry of the energy gap was different 
from that in
conventional low-$T_c$ materials. In particular, the energy gap was
found to vanish along certain directions in momentum space. These
so-called ``nodes'' serve as a conduit
for extreme low-energy QP excitations.
One of the key early experiments providing evidence for the
existence of gap nodes was microwave measurements by Hardy {\it et al.}
\cite{Hardy:93} of the in-plane penetration depth change
$\Delta \lambda_{ab} \! = \! \lambda_{ab}(T) \! - \! \lambda_{ab}(1.35$~K) 
in the Meissner state of high-purity YBa$_2$Cu$_3$O$_{6.95}$. 
In this phase, magnetic field is partially screened from the interior by
``supercurrents'' circulating around the sample perimeter. These
supercurrents constitute the response of the superconductor to the
applied field. 
The penetration depth $\lambda$ is the characteristic
length scale over which the field decays
in from the surface, and the quantity $\lambda^{-2}$ 
is proportional to the density of Cooper pairs, 
{\it i.e.} ``superfluid density'', $n_s$.
Because thermal energy can excite QPs, $\lambda^{-2}$ decreases
with increasing $T$. In a conventional superconductor,
this temperature dependence is typically weak at low $T$ because  
the isotropic energy gap exponentially cuts off the QP excitations as
$T \! \rightarrow \! 0$~K. In Ref.~\cite{Hardy:93}, however, 
$\lambda_{ab}^{-2}$ was found to decrease sharply upon 
raising the temperature above 
1.35~K---the lowest temperature reached in the experiment. 
This suggested that the minimum gap size was
very small. Moreover, at low temperatures $\Delta \lambda_{ab}$ was observed to be
proportional to $T$, which is characteristic of a superconducting order 
parameter which has $d_{x^2-y^2}$-wave symmetry (rather than $s$-wave
symmetry as in conventional superconductors).
Later the same behavior for $\lambda_{ab}$ 
was observed in the vortex state of YBa$_2$Cu$_2$O$_{6.95}$,
using the muon spin rotation
($\mu$SR) technique \cite{Sonier:94}. In the vortex state, magnetic
field penetrates the sample in the form of quantized flux lines, called
``vortices'', which usually arrange themselves into a periodic array. 
Screening currents circulate around the individual flux
lines, so that here $\lambda_{ab}$ is associated with the decay of the
field outside the vortex cores. 

Recently, researchers have been wondering whether the story of the gap
symmetry in the high-$T_c$ materials is complete. Some experiments
performed at high magnetic fields in the vortex state show features 
which are seemingly inconsistent with a pure $d_{x^2-y^2}$-wave order 
parameter. For instance, scanning tunneling spectroscopy (STS) measurements
in YBa$_2$Cu$_3$O$_{7-\delta}$ (YBCO) suggest the existence
of localized QP states in the vortex cores \cite{Maggio:95}, which can only 
exist if there is an energy gap over the entire Fermi surface.
Equally intriguing have been reports of an unusual
plateau-like feature in the field dependence of the thermal
conductivity $\kappa(H)$ in Bi$_2$Sr$_2$CaCu$_2$O$_8$ (BSCCO) 
\cite{Krishana:97,Aubin:98,Ando:99}, and
underdoped YBa$_2$Cu$_3$O$_{6.63}$ \cite{Ong:99}. 
A field-induced transition from
a $d_{x^2-y^2}$-wave state to a fully gapped state, such as 
$d_{x^2-y^2} + i d_{xy}$ or $d_{x^2-y^2} + i s$, has been offered
as a possible explanation for these results \cite{Laughlin:98,Franz:98}.
However, subsequent measurements of $\kappa(H)$ have established that
the plateau is hysteretic \cite{Aubin:98} and is
not observed in all samples \cite{Ando:99,Ong:99}. These observations 
suggest that the vortex lattice (VL) and/or impurities are partially
responsible for this feature.

Recent developments in spectrometer design at the TRI-University Meson
Facility (TRIUMF) in Vancouver, Canada have made it possible to
extend the application of the $\mu$SR technique to magnetic fields
as high as 8~T. This has allowed us to measure $\lambda$ at fields where the
anomalies in the STS and $\kappa(H)$ measurements have been
observed. In a
$\mu$SR experiment the inhomogeneous magnetic field associated with the
VL is sensitively measured by implanting muons into the
sample, where their spins precess with a frequency which is directly 
proportional to the local field. The muon-spin precession signal which
is obtained by detecting a muon's positron-decay pattern, contains
the precession frequencies from all the muon stopping sites. 
The line width of the corresponding field distribution is 
roughly proportional to $\lambda^{-2}$.

Our measurements were performed on {\em detwinned} single crystals of 
YBa$_2$Cu$_3$O$_{6.95}$ ($T_c \! = \! 93.2$~K), where a linear 
$T$-dependence has been previously observed at low $H$
\cite{Sonier:94,Sonier:97B}. To generate a well ordered VL, the sample 
was cooled in the applied field. Neutron scattering 
measurements on a {\em detwinned} crystal of YBCO over
the field range $H \! = \! 0.2$-4~T, show that at low $T$ the vortices 
arrange themselves in a hexagonal lattice, distorted by $\hat{a}$-$\hat{b}$
anistropy \cite{Johnson:99}. This is consistent with recent theoretical 
calculations which show that in a $d_{x^2-y^2}$-wave superconductor 
the VL is hexagonal at reduced fields $H/H_{c2} \! < \! 0.15$
\cite{Ichioka:99}.
 
An example of the measured muon spin precession signal 
is shown in Fig.~1(a). Although the time spectrum itself is not 
very revealing to the eye, a fast Fourier transform 
(FFT) can be performed, as in Fig.~1(b), to illustrate
the distribution of precession frequencies (or local fields) in the sample.
We stress that the frequency spectrum in Fig.~1(b) is only an 
{\em approximation} of the actual internal field distribution, because
the finite time range and the ``apodization'' needed to eliminate
``ringing'' in the FFT artificially broadens the output spectrum
\cite{Fleming:82}. We have previously shown that a sharp field
distribution exists in high quality YBa$_2$Cu$_3$O$_{6.95}$
crystals, with the basic features slightly smeared out by a 
small random displacement of the vortex positions and a
small nuclear dipolar contribution \cite{Sonier:97B}. 
Despite the shortcomings of the FFT procedure, the basic features 
of a hexagonal VL are recognizable in the observed aymmetric line 
shape---namely, the minimum field at the center of the triangle formed
by three adjacent vortices is close to the cusp 
associated with the Van Hove singularity produced by the 
field at the center position between nearest-neighbor vortices.  
The observation of a high-field ``tail'',
related to the field in the vicinity of the vortex cores, is a
signature of a well-ordered VL. 
The inset of Fig.~1(b) shows the
field dependence of the skewness parameter, 
$\alpha \! = \! \langle (\Delta B)^3 \rangle^{1/3} /
\langle (\Delta B)^2 \rangle^{1/2}$ [where
$\langle (\Delta B)^n \rangle \! = \! \langle 
(B \! - \! \langle B \rangle)^n \rangle$], which characterizes
the symmetry of the field distribution. The decrease in $\alpha$
with increasing $H$ is predicted to occur in both $s$- and
$d_{x^2-y^2}$-wave superconductors \cite{Ichioka:99}, and 
does not imply a change in VL geometry. The qualitative agreement
between the behavior of $\alpha(H)$ extrapolated to $T \! = \! 0$~K 
and at $T \! = \! 50$~K, suggests that the VL remains hexagonal
over this temperature range. 

The muon-spin precession signals were fit assuming an analytical
Ginzburg-Landau (GL) model for the internal field profile associated
with the VL \cite{Yaouanc:97}
\begin{equation}
B(r) = B_0 (1-b^4)\sum_{ {\bf G}}
{ e^{-i {\bf G} \cdot r}
\,\, u \, K_1(u)
\over
\lambda_{ab}^2 G^2},
\label{eq:GL}
\end{equation}
where $u^2 \! = \! 2 \xi_{ab}^2 G^2 (1+b^4)[1-2b(1-b)^2]$,
$B_0$ is the average internal field, ${\bf G}$ are the reciprocal
lattice vectors, $\xi_{ab}$ is the in-plane GL coherence length and
$K_1(u)$ is a modified Bessel function. The fitting procedure is described in 
detail elsewhere \cite{Sonier:97}. We have verified that 
the qualitative results described in this Letter are robust 
with respect to the theoretical model assumed for $B(r)$.
The solid curves in Fig.~1(a) are an example of a fit of 
the time spectrum 
to a theoretical complex muon polarization function 
$\tilde{P}(t) \! = \! P_x(t) \! + \! P_y(t)$, assuming the field
profile in Eq.~(\ref{eq:GL}) 
and a Gaussian distribution of fields for the small 
background signal barely visible in the FFT at 813.4~MHz. 

Figure~2 shows the temperature dependence of $\lambda_{ab}^{-2}$
at $H \! = \! 0.5$, 4 and 6~T. The solid curve in Fig.~2,
which agrees very well with the $H \! = \! 0.5$~T data at low $T$, represents
the microwave measurements of $\Delta \lambda_{ab}$
at zero DC field \cite{Hardy:93}, converted to $\lambda_{ab}^{-2}$
using the $\mu$SR value of $\lambda_{ab}$ at $T \! = \! 1.35$~K.
At the higher fields the linear $T$-dependence at low temperatures
vanishes and the magnitude of $\lambda_{ab}^{-2} \! \propto \! n_s$
is reduced. This occurs below $T \! \approx \! 25$~K 
and 35~K at $H \! = \! 4$~T and 6~T,
respectively. While the nature of the QP excitations at 
low $T$ is strongly influenced by the applied field,
at intermediate temperatures (35~K~$\! < \! T \! < \! 60$~K)
the field has a negligible effect. At higher $T$ the VL
melts at a field-dependent temperature $T_{\rm m} (H)$, as shown
by magnetization measurements on similar crystals \cite{Billon:97}.
In the melted region it is not straightforward 
to extract $\lambda_{ab}$. The observed decrease in $\lambda_{ab}^{-2}$
above $T_m(H)$ arises from a loss of asymmetry in the
measured field distribution. 
We now proceed to discuss the possible sources of the
field-induced reduction of $\lambda_{ab}^{-2}$ at low $T$.
    
The application of a magnetic field results in a
shift of the QP energy levels due to the superflow \cite{Tinkham}. 
In a $d_{x^2-y^2}$-wave superconductor the low-lying QP levels 
near the nodes are shifted below the Fermi surface,
resulting in a QP current which flows in opposite direction to the 
superflow. This leads to a {\em nonlinear} relationship
between the supercurrent density $J_s$ and the superfluid velocity
$v_s$. The net result is a weakened supercurrent response,
leading to an increased penetration of the field. In the
Meissner state, Yip and Sauls \cite{Yip:92} predicted that
nonlinear effects in a $d_{x^2-y^2}$-wave superconductor
give rise to a linear $H$-dependence
for $\lambda$ at $T \! = \! 0$~K. 
Although recent measurements of $\lambda (H)$ 
\cite{Bidinosti:99,Carrington:99} show evidence for some
sort of nonlinear Meissner effect, they do not agree 
with the Yip and Sauls prediction, and do not show a
saturation at low $T$ as observed here.   
   
According to Amin, Affleck and Franz \cite{Amin:98}, a
more dominant contribution to the field dependence of $\lambda_{ab}$
measured by $\mu$SR, comes from the {\em nonlocality} of 
the supercurrent response in the vicinity of the gap nodes.
In a clean $d_{x^2-y^2}$-wave superconductor the coherence
length $\xi_0$, which is inversely proportional to the gap size,
diverges at the nodes. Thus, near these regions of the Fermi 
surface, $\xi_0 \! > \! \lambda$,  
and $J_s$ at a point $r$ is obtained by averaging 
the field over a surrounding region of radius $\xi_0$. 
For a non-uniform field, nonlocal effects weaken the 
supercurrent response. In Ref.~\cite{Amin:98}, the 
{\em effective} $\lambda_{ab}(H)$ measured by $\mu$SR
was calculated for the combination of nonlocal and nonlinear 
effects in a pure $d_{x^2-y^2}$-wave superconductor in the vortex
state at $T \! = \! 0$~K. The result is that $\lambda_{ab}(H)$
is a nonlinear function of magnetic field, predominantly due to 
a modification of the field distribution by nonlocal effects.   
In Fig.~3 we show that our measurements of 
$\lambda_{ab}(H)$ agree extremely well with this prediction. 
Although no prediction for the temperature dependence
of $\lambda_{ab}$ appears in Ref.~\cite{Amin:98}, recent
calculations show a continuous transition as a function of $T$ to 
the $T$-independent region \cite{Amin:99}. 
We remark that nonlocal effects are not
expected to play a role in the high-field
anomalies observed in the STS \cite{Maggio:95}
and thermal conductivity \cite{Krishana:97} experiments.
 
Another possible explanation for the saturation of
$\lambda_{ab}^{-2}$ at low $T$ is that a field-induced transition
to a fully gapped state has taken place which freezes
out the QP excitations. This does not necessarily lead to a change 
in VL geometry---since as noted earlier, 
the VL is predicted to be hexagonal at low $H/H_{c2}$ irrespective of 
the pairing symmetry \cite{Ichioka:99}. Furthermore,
at low $T$ the vortices in YBCO are strongly pinned 
and probably reluctant to redistribute themselves.
However, if a small imaginary component ($s$- or $d_{xy}$-wave)
is induced by the field at low $T$, naively one would expect
$\lambda_{ab}^{-2} (T)$ to saturate at a value equivalent to
$\lambda_{ab}^{-2} (T \! \rightarrow \! 0$~K$)$ at $H \! = \! 0.5$~T,
which is clearly not what we observe.
Recently, Yasui and Kita \cite{Yasui:99} have shown by self-consistently
solving the Bogoliubov-de-Gennes (BdG) equations, that the mixing of an
$s$- or $d_{xy}$-wave component, if it exists,
must gradually develop as a function of field in the vortex state.
Although this is not inconsistent with our $\mu$SR measurements, 
the high-field plateau observed in the $\kappa(H)$ measurements \cite{Krishana:97}
indicates a sharp field-induced transition of unknown origin.
In Ref.~\cite{Yasui:99} it is also shown
that the double-peak structure observed near 
zero bias in the STS experiment of Maggio-Aprile {\it et al.} \cite{Maggio:95} 
can originate from low-energy QP hopping between the
vortex cores of a pure $d_{x^2-y^2}$-wave superconductor---negating the 
need for the large $d_{xy}$ component suggested
by Franz and Te\u{s}anovi\'{c} \cite{Franz:98} in an earlier treatment
of an isolated vortex using the BdG formalism.
Thus, current theories can explain both the STS and $\mu$SR
anomalies at high field in terms of a pure $d_{x^2-y^2}$-wave superconductor.
We note that while the peak structure was not observed in STS
measurements of the vortex cores in BSCCO by Renner {\it et al.}
\cite{Renner:98}, a recent STS study by Davis {\it et al.} \cite{Davis:99}
shows evidence for QP core states in this material.

Finally, Fig.~4 shows the field dependence of the in-plane coherence
length $\xi_{ab}$ from the fits to the $\mu$SR spectra, 
extrapolated to $T \! = \! 0$~K.
Previously, we established that $\xi_{ab}$ is nearly proportional
to the size of the vortex cores $r_0$, and that $r_0$ expands at low
fields \cite{Sonier:99}. The high-field measurements clearly show that
the core size saturates with $\xi_{ab} \! \approx \! 18.5$~\AA~
when the density of vortices ($\sim \! H$) becomes large. The
expansion of $r_0$ at low fields is qualitatively consistent with
the predictions from a recent quasiclassical theoretical 
study of the vortex structure in pure $s$- and $d_{x^2-y^2}$-wave 
superconductors \cite{Ichioka:99}.  

In summary, we observe a field-induced saturation in the
temperature dependence of $\lambda_{ab}$ at low $T$ measured 
by $\mu$SR. The behavior of $\lambda_{ab}^{-2} (T)$ appears inconsistent with
a transition to a mixed $d_{x^2-y^2} \! + \! id_{xy}$ ($d_{x^2-y^2} \! + \! is$) 
state. The measured field dependence of $\lambda_{ab}$ is in agreement 
with the theory of nonlocal and nonlinear effects
in the vortex state of a pure $d_{x^2-y^2}$-wave superconductor.
 
We are especially grateful to Syd Kreitzman, Mel Good, Donald
Arseneau and Bassam Hitti for developing the high-field $\mu$SR
apparatus. We thank Alexander Balatsky, Mohammad~Amin, Ian Affleck,
Ilya Vekhter, Chris Bidinosti 
and Peter~Hirschfeld for informative discussions.
This work is supported by the Natural Sciences and Engineering Research
Council of Canada, the Canadian Institute for
Advanced Research, the New Energy Development Organization of
Japan and at Los Alamos by the US Department of Energy.

\newpage
\begin{center} 
FIGURE CAPTIONS 
\end{center} 
 
Figure 1. (a) The muon spin precession signal at $T \! = \! 20$~K and
$H \! = \! 6.0$~T displayed in a reference frame rotating at about 3~MHz
below the Larmor precession frequency of a free muon in the external field
[Note: $A$ is the maximum precession amplitude].  
(b) The FFT of (a) using a Gaussian apodization with a 3~$\mu$s time constant.
Inset: field dependence of the skewness parameter $\alpha$ extrapolated to
$T \! = \! 0$~K [solid circles] and at $T \! = \! 50$~K [open circles].\\

Figure 2. Temperature dependence of $\lambda_{ab}^{-2}$ at 
$H \! = \! 0.5$~T , 4~T and 6~T. The solid curve represents the zero-field 
microwave measurements of $\Delta \lambda_{ab}(T)$ in Ref.~\cite{Hardy:93}.\\

Figure 3. Magnetic field dependence of $\lambda_{ab}$ in
YBa$_2$Cu$_3$O$_{6.95}$ extrapolated to $T \! = \! 0$~K 
[open circles] and the predicted behavior at $T \! = \! 0$~K
from Ref.~\cite{Amin:98} for the combination of nonlinear 
and nonlocal effects in the
vortex state of a $d_{x^2-y^2}$-wave superconductor [solid diamonds].\\

Figure 4. Magnetic field dependence of $\xi_{ab}$ in 
YBa$_2$Cu$_3$O$_{6.95}$ extrapolated to $T \! = \! 0$~K.\\  


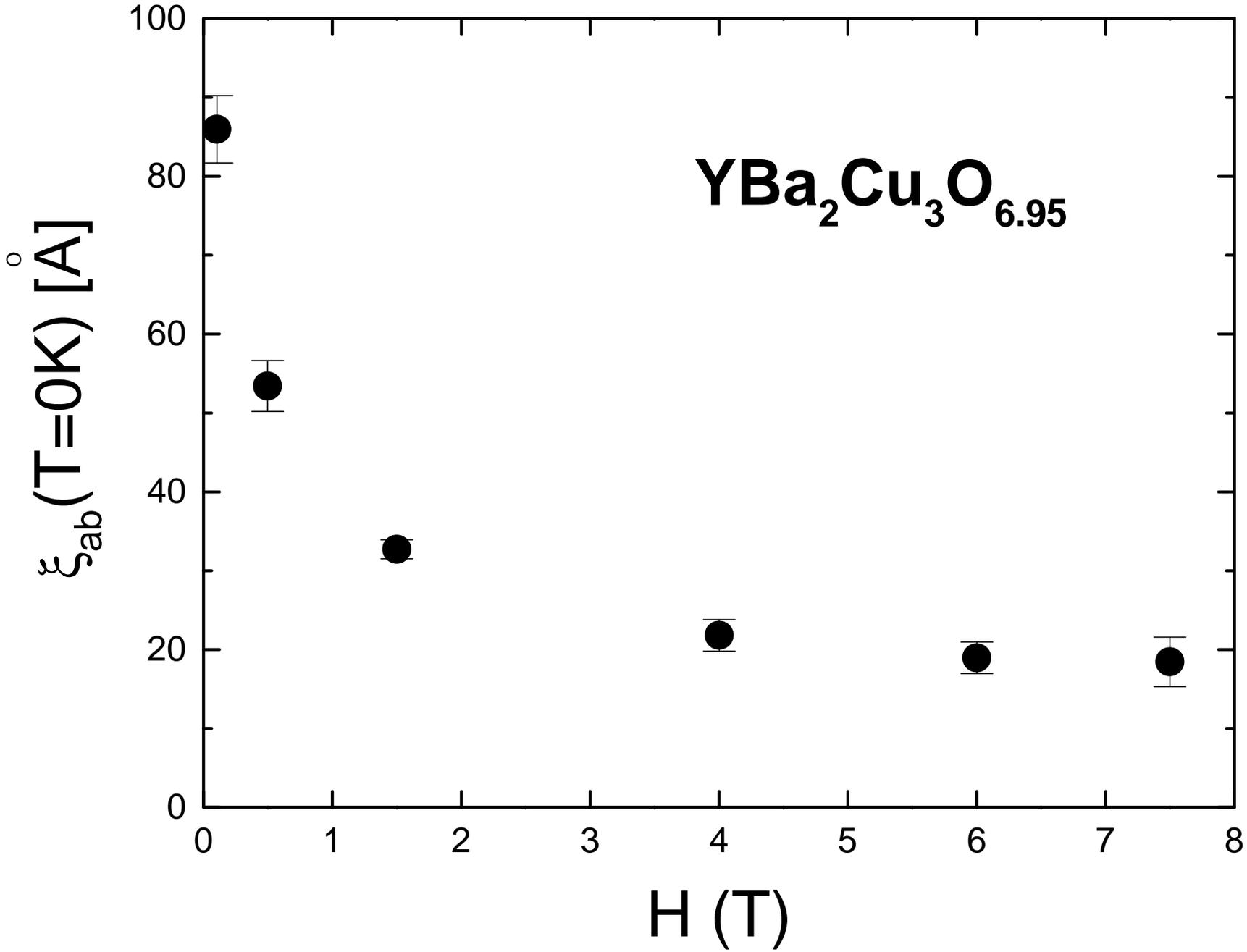
\end{document}